\documentclass[12pt,preprint]{aastex}

\usepackage{multirow}
\usepackage{hhline}

\def\ltsima{$\; \buildrel < \over \sim \;$}
\def\gtsima{$\; \buildrel > \over \sim \;$}
\def\lsim{\lower.5ex\hbox{\ltsima}}
\def\gsim{\lower.5ex\hbox{\gtsima}}
\def\lapp{\ifmmode\stackrel{<}{_{\sim}}\else$\stackrel{<}{_{\sim}}$\fi}
\def\gapp{\ifmmode\stackrel{>}{_{\sim}}\else$\stackrel{<}{_{\sim}}$\fi}

\def\U{F390W}
\def\V{F606W}
\def\I{F814W}
\def\H{F656N}
\def\B{F435W}
\def\R{F625W}
\def\Ha{F658N}
\def\nUV{F255W}
\def\fUV{F170W}
\def\VV{F555W}
\def\UU{F336W}
\def\halpha{H$\alpha$}
\newcommand{\at}{\makeatletter @\makeatother}

\def\igr{IGR J1824-24525}

\newdimen\minuswidth    

\setbox0=\hbox{$-$}

\minuswidth=\wd0

\catcode`@=\active

\def@{\kern\minuswidth}

\setbox0=\hbox{\rm0}

\shorttitle{The optical counterpart to IGR J18245-2452}

\shortauthors{C. Pallanca}

\begin{document} 

\title{The optical counterpart to the X-ray transient  \igr\ in the globular cluster M28\footnote{Based on observations
    collected with the NASA/ESA HST (Prop. 19835), obtained at the
    Space Telescope Science Institute, which is operated by AURA,
    Inc., under NASA contract NAS5-26555.}}

\author{
C. Pallanca\altaffilmark{2},
E. Dalessandro\altaffilmark{2},
F.R. Ferraro\altaffilmark{2},
B. Lanzoni\altaffilmark{2} and
G. Beccari\altaffilmark{3}.
}
\affil{\altaffilmark{2} Dipartimento di Fisica e Astronomia, Universit\`a degli
  Studi di Bologna, Viale Berti Pichat 6/2, I--40127 Bologna, Italy; cristina.pallanca3\at unibo.it}
\affil{\altaffilmark{3} European Southern Observatory,
  Karl-Schwarzschild-Strasse 2, 85748 Garching bei M\"unchen, Germany}

\date{24 Jun, 2013}

\begin{abstract}
We report on the identification of the optical counterpart to the
recently detected INTEGRAL transient \igr\ in the Galactic globular
cluster M28.  From the analysis of a multi epoch HST dataset we have identified a
strongly variable star positionally coincident with the radio and
Chandra X-ray sources associated to the INTEGRAL transient.  The star
has been detected during both a quiescent and an outburst state.  In
the former case it appears as a faint, unperturbed main sequence star, while in
the latter state it is about two magnitudes brighter and slightly bluer
than main sequence stars.  We also detected \halpha\ excess during the outburst
state, suggestive of active accretion processes by the neutron star.
\end{abstract} 


\section{INTRODUCTION}\label{intro}

The high stellar densities and the frequent dynamical interactions
occurring in globular cluster (GC) cores are expected to significantly
affect the formation and the evolution of exotic populations, such as
low-mass X-ray binaries (LMXBs), cataclysmic variables, millisecond
pulsars (MSPs), and blue straggler stars \citep[e.g.][]{bailyn95,
  verb97,  grindlay01, pooley03, fe09_m30}.  In fact, these objects are thought to
result from the evolution of various kinds of binary systems
originated and/or hardened by stellar interactions
\citep[e.g.][]{clark75, hillsday76, bailyn92, ivanova08}, and are
therefore considered as powerful diagnostics of GC dynamical evolution
\citep[e.g.][]{fe95, goodmanhut89,hut92, meylanheggie97, pooley03,
  fregeau08, fe12}. However, many open questions still remain about
their formation and evolutionary paths.

When a close binary system contains a compact object, mass transfer
processes can take place. The streaming gas, its impact on the compact
star, or the presence of an accretion disk can produce significant
X-ray and UV radiation, together with emission lines (such as the
\halpha) or rapid luminosity variations.  The first evidence of
interacting binaries in Galactic GCs was indeed obtained through the
discovery of X-ray sources.  In particular, LMXBs are thought to be
binary systems with an accreting neutron star (NS) and are
characterized by X-ray luminosities larger than $\sim 10^{35}$ erg
s$^{-1}$. Their final stage is thought to be a binary system
containing a very fast NS (a MSP), spun up through mass accretion from
the evolving companion.   Moreover, during their life  some LMXBs,  usually called X-ray transients \citep{white84}, show a
few outbursts and during the quiescent state their millisecond pulsation can become detectable \citep[][]{chak98}.
 
The identification of the optical counterparts is a fundamental step
for characterizing these exotic binary systems, both in quiescent and
in outburst state, and for clarifying their formation and evolutionary
processes \citep[][]{testa12}.  Determining the nature and the properties of the companion
(which dominates the optical emission in the quiescent state) is also very
useful to tightly constrain the orbital parameters of the system
\citep[e.g.][]{d'avanzo09, engel12}.  In the case of GCs, it also
represents a crucial tool for quantifying the occurrence of dynamical
interactions, understanding the effects of crowded stellar
environments on the evolution of binaries, and determining the shape
of the GC potential well \citep[e.g.][]{phinney92, bellazz95, 
  possenti03, fe03}.

M28 (NGC 6626) is a Galactic GC with intermediate central density
\citep[$\log \rho_0=4.9$ in units of ${\rm M}_\odot$pc$^{-3}$;][]{
  pryor93} and relatively high metallicity
\cite[{[Fe/H]}=-1.32,][ 2010
  version]{harris96} located at $\sim 6.8$ kpc from
Earth, in the direction of the Galactic centre \citep[][]{harris96}.  It is the first Galactic GC where a MSP was
discovered \citep{lyne87} and it is currently known to host the third
largest population of pulsars among all GCs \citep{begin06,
  bogd11}\footnote{For the complete list of pulsars in Galactic GCsunambiguously
  see the web site http://www.naic.edu/$\sim$pfreire/GCpsr.html}.  A
total of 46 X-ray sources, of which 12 lie within one core radius
\citep[$r_c=14.4\arcsec$;][]{harris96} from the centre, has been
detected with Chandra \citep{becker03}.

During the observations of the Galactic center performed on 2013 March
28 with INTEGRAL, a new hard X-ray transient (IGR J18245-2452) has
been revealed in the direction of M28 (Eckert et al., ATel \#4925).
Subsequent observations with SWIFT/XRT confirmed the detection of the
transient source and its location within the core of the cluster, at
$\alpha_{2000}=18^{\rm h} 24^{\rm m} 32.20^{\rm s}$ and
$\delta_{2000}=-24^\circ 52' 05.5\arcsec$, with error radius
$3.5\arcsec$ (at 90\% confidence; Heinke et al., Atel \#4927; Romano
et al., ATel \#4929).  SWIFT/XRT time-resolved spectroscopy performed
on 2013 April 7 have revealed a thermal spectrum with a ``cooling
tail'', unambiguously identifying the burst as thermonuclear and
suggesting that the source is a low-luminosity LMXB in the hard state,
where a NS is accreting matter from a companion (Linares,
Atel. \#4960; see also Serino et al., Atel \#4961).  A radio follow-up
has been performed with ATCA on 2013 April 5, for a total of 6 hours,
at two different frequencies (9 and 5.5 Hz). A single source has been
identified at $\alpha_{2000}=18^{\rm h} 24^{\rm m} 32.51^{\rm s}$ and
$\delta_{2000}=-24^\circ 52' 07.9\arcsec$, with a 90\% confidence
error of $0.5\arcsec$ (Pavan et al., Atel \#4981).  This position is
only marginally consistent with that derived from the SWIFT/XRT data,
but the detected strong variability (reaching up to 2.5 times the mean
flux density during the first 90 minutes of observations) suggests a
possible association with the X-ray  transient. Its position well
corresponds  to the location of the X-ray source \#23 identified by
\citet{becker03} from Chandra observations and associated to \igr\ by
Homan et al. (ATel \#5045).

Here we report on the identification of the optical counterpart to
\igr, obtained from the analysis of high resolution HST data acquired
with the WFPC2, WFC3 and ACS/WFC in three different epochs (see also
Pallanca et al. Atel\#5003, and Cohn et al. Atel \#5031).  In Section
\ref{obs} we describe the dataset and the data analysis procedure.
The properties of the optical counterpart to \igr\ are presented in
Section \ref{identification} and discussed in Section
\ref{conclusions}.

\section{OBSERVATIONS AND  DATA ANALYSIS}\label{obs}
For this work we adopted the same catalog used to identify the
companion to PSR J1824-2452H and fully described in
\citet{pallanca10}.  In order to unveil luminosity variations
among different epochs, two additional sets of HST data acquired with
the WFPC2 and the ACS have been analyzed. In particular, because we
were interested only in the GC core, we limited the analysis to the
Planetary Camera (PC) of the WFPC2 and  CHIP2 of the ACS/WFC
mosaic.  The available samples have been acquired through various
filters, at three different epochs (see Table \ref{dataset}): the
WFPC2 dataset was collected on 2009, April 7 (epoch 1, hereafter EP1),
WFC3 observations were performed on 2009, August 9 (epoch 2, EP2) and
the ACS data-set was acquired on 2010, April 26 (epoch 3, EP3).

\begin{table}
\centering\begin{tabular}{ c  c  l  c  l  c  l }
\hline
\hline
Epoch & Date & Instrument & Filter & ${\rm t}_{{\rm exp}}$ [s] & State & Proposal ID/PI\\
\hline
\multirow{4}{*}{{ \small EP1}}   & \multirow{4}{*}{{ \small 2009/04/07}}   & \multirow{4}{*}{{ \small WFPC2/PC}}    & { \scriptsize F170W} & { \scriptsize $2 \times1700$}  &\multirow{4}{*}{{ \small Q}}  & \multirow{4}{*}{{ \small GO11975/Ferraro}}\\
\multirow{4}{*} {}          & \multirow{4}{*} {}                      &  \multirow{4}{*} {}                       & { \scriptsize F255W} &  { \scriptsize $3 \times 1200$}   &  \multirow{4}{*}{}   & \multirow{4}{*}{}\\
\multirow{4}{*} {}          & \multirow{4}{*} {}                      &  \multirow{4}{*} {}                       & { \scriptsize F336W} &  { \scriptsize $3 \times 800$}     &  \multirow{4}{*}{}   & \multirow{4}{*}{}\\
\multirow{4}{*} {}          & \multirow{4}{*} {}                      &  \multirow{4}{*} {}                       & { \scriptsize F555W} &  { \scriptsize $2 \times 80$}       &  \multirow{4}{*}{}   & \multirow{4}{*}{}\\
\hline
\multirow{5}{*} {{ \small EP2}}  & \multirow{5}{*} {{ \small 2009/08/09}}  &  \multirow{5}{*} {{ \small WFC3/UVIS}} & { \scriptsize F390W} & { \scriptsize $5\times 850+1\times800$}    &  \multirow{5}{*}{{ \small B}} &  \multirow{5}{*}{{ \small GO11615/Ferraro}}\\
\multirow{5}{*} {}          & \multirow{5}{*} {}                      &  \multirow{5}{*} {}                       &{ \scriptsize  F606W} & { \scriptsize  $7 \times 200$}     &  \multirow{5}{*}{}    & \multirow{5}{*}{}\\
\multirow{5}{*} {}          & \multirow{5}{*} {}                      &  \multirow{5}{*} {}                       & { \scriptsize F814W} & { \scriptsize $7 \times 200$}      &  \multirow{5}{*}{}    & \multirow{5}{*}{}\\
\multirow{5}{*} {}          & \multirow{5}{*} {}                      &  \multirow{5}{*} {}                       & { \scriptsize F656N}  &  { \scriptsize $2\times1100+1\times1070$}&  \multirow{5}{*}{}    & \multirow{5}{*}{}\\
\multirow{5}{*} {}          & \multirow{5}{*} {}                      &  \multirow{5}{*} {}                       &  \multirow{5}{*} {}      &  { \scriptsize $3\times1020+1\times935$}&  \multirow{5}{*}{}    & \multirow{5}{*}{}\\
\hline
\multirow{3}{*}{{ \small EP3}}   & \multirow{3}{*}{{ \small 2010/04/26}}   & \multirow{3}{*}{{ \small ACS/WFC}}      & { \scriptsize F435W}  & { \scriptsize $4 \times 464$}      &\multirow{3}{*}{{ \small Q}}  & \multirow{3}{*}{{ \small GO11340/Grindlay}}\\ 
\multirow{3}{*} {}          & \multirow{3}{*} {}                      &  \multirow{3}{*} {}                       & { \scriptsize F625W}  & { \scriptsize $4 \times 60$}        &  \multirow{3}{*}{}   & \multirow{3}{*}{}\\
\multirow{3}{*} {}          & \multirow{3}{*} {}                      &  \multirow{3}{*} {}                       & { \scriptsize F658N}   & { \scriptsize $6\times724 + 3\times717$}   &  \multirow{3}{*}{}   & \multirow{3}{*}{}\\
\hline
\end{tabular}
\caption{Summary of the multi-epoch data-sets used in this work.  The quiescent and outburst state (see Section \ref{identification}) are marked by letters Q and B, respectively.}
\label{dataset}
\end{table}

The data reduction procedure for the ACS sample has been performed on
the CTE-corrected (flc) images, once corrected for Pixel-Area-Map
(PAM) by using standard IRAF procedures.  The photometric analysis has
been carried out by using the DAOPHOT package \citep{stetson87}. For
each image we modeled the point spread function (PSF) by using a large
number ($\sim 100$) of bright and nearly isolated stars.  Then, all
\B\ and \R\ images have been combined with MONTAGE2 and used to
produce a master frame on which we optimized a master list of stars.
Finally we performed the PSF fitting on this master list by using the DAOPHOT
packages ALLSTAR and ALLFRAME \citep{stetson87, stetson94}.  A similar
procedure has been adopted to reduce the flat-fielded (flt) WFPC2
images.

Since the ACS images heavily suffer from geometric distortions within
the field of view, we corrected the instrumental positions of stars by
applying the equations reported by \citet{sirianni05}.  We then
placed, through cross-correlation, the ACS and the WFPC2 data sets on
the same astrometric system of the WFC3 sample, for which the
astrometric solution has an accuracy of $\sim0.2''$ in both right
ascension and declination \citep{pallanca10}.

Finally, the instrumental magnitudes have been calibrated to the
VEGAMAG system by using the photometric zero-points reported on the
instrument web
pages\footnote{www.stsci.edu/hst/acs/analysis/zeropoints/zpt.py and
  www.stsci.edu/documents/dhb/web/c32$\_$wfpc2 dataanal.fm1.html for
  ACS and WFPC2, respectively} and the procedure described in
\citet{holtzman95} and \citet{sirianni05} for WFPC2 and ACS,
respectively.

\section{THE OPTICAL COUNTERPART TO IGR J18245-2452}\label{identification}
\begin{figure*}
\includegraphics[width=150mm]{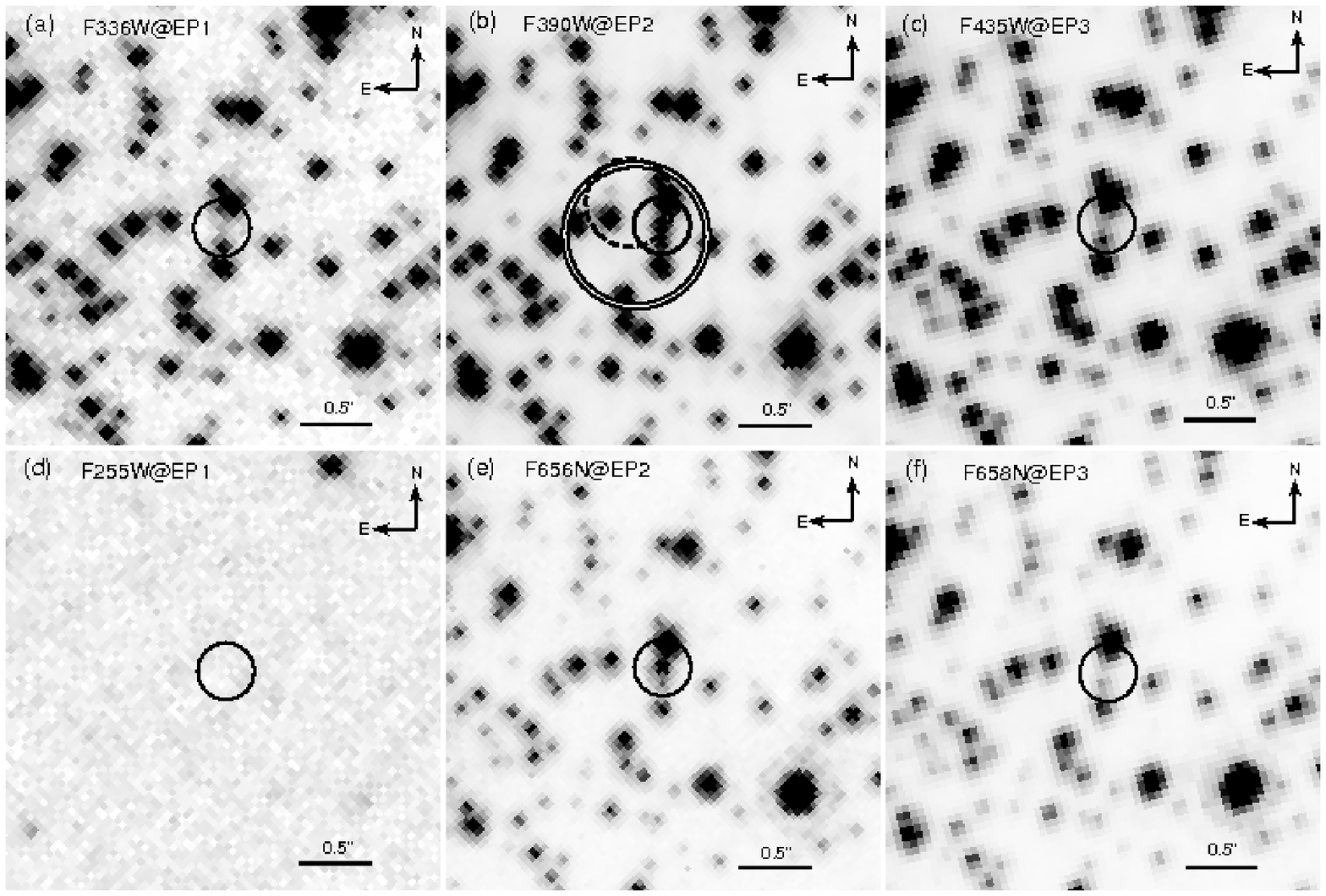}
\caption{HST images of the optical counterpart (solid circle) to
  \igr. The filters and epochs of observation are labelled in each
  panel (see Table \ref{dataset} for more details).  Clearly, the
  source is in a quiescent state in EP1 and EP3 (leftmost and
  rightmost panels), while it has been caught in outburst during EP2
  (central panels). In panel (b) the double and dashed circles mark,
  respectively, the position of the variable ATCA source detected by
  Pavan et al. (ATel \#4981) and the Chandra X-ray source \# 23
  \citep{becker03}, with the radii corresponding to the quoted
  astrometric uncertainties.}
\label{map}
\end{figure*}

During a systematic study of the GC M28 aimed at searching for the
companion stars to binary MSPs, we found a peculiar object (see Figure
\ref{map}) located at $\alpha_{2000}=18^{\rm h}24^{\rm m}32.50^{\rm
  s}$ and $\delta_{2000}=-24^\circ52'07.8''$, in very good agreement
with the position of the X-ray source \#23 reported by
\citet{becker03} and of the variable ATCA radio source discussed by Pavan
et al.  (Atel \#4981).

\begin{figure*}
\includegraphics[width=150mm]{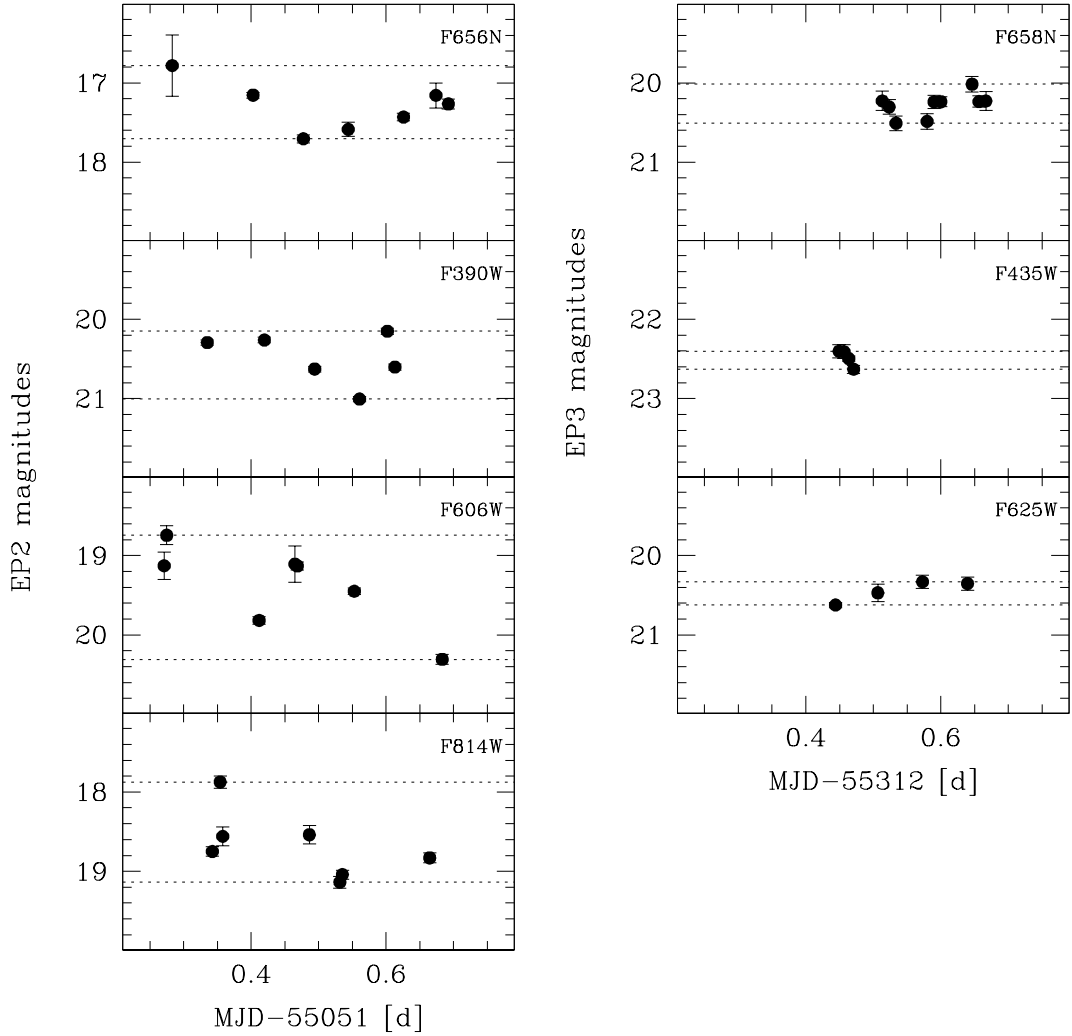}
  \caption{Light curves of the optical counterpart to \igr\ during the outburst state (left panels) and the  quiescent state (right panels). The dotted lines mark the maximum  range of variability detected by each set of observations. Photometric errors are reported, but in most
    cases are smaller than the point size.}
\label{time}
\end{figure*}

\begin{figure*}
\includegraphics[width=150mm]{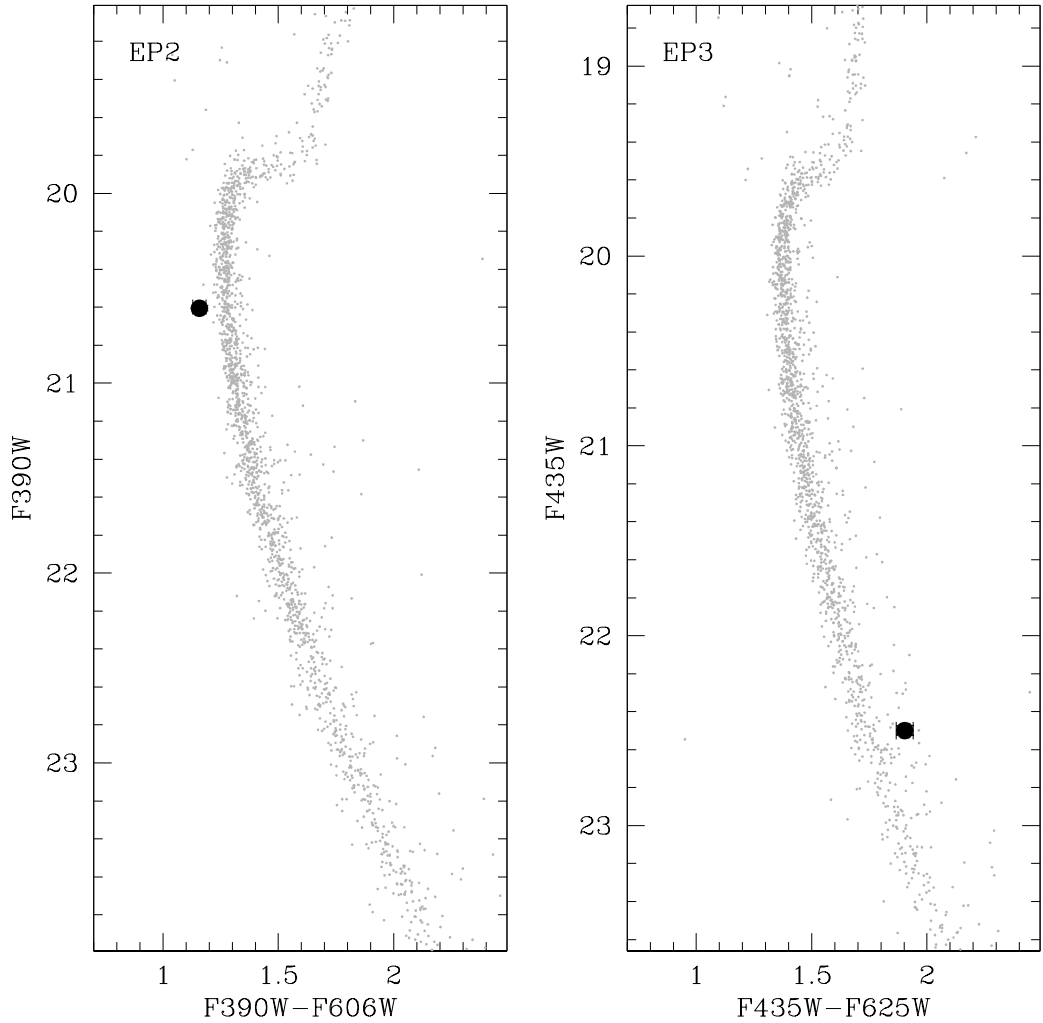}
  \caption{Color magnitude diagrams obtained during outburst epoch (left panel) and quiescence epoch (right panel) for all  stars (gray points) located  within 10\arcsec\ from \igr.  The location of the optical counterpart to \igr, as obtained by averaging the observed light curves (see Figure \ref{time} and footnote 3), is shown as a large solid circle.}
\label{cmd}
\end{figure*}

In EP2 this star showed a strong and irregular variability in each
filter, on a timescale of $\sim 10$ hours (Figure \ref{time}). Based
on the mean magnitudes\footnote{It is important to note that, given
  the  variability and an undersampled time coverage, the mean
  magnitudes  (and hence the  colors) derived here could not exactly correspond 
  to the true average luminosities of the star over the entire variability period.}  (\U$=20.61\pm0.01$, \V$ =19.45\pm0.02$, \I$
=18.83\pm0.03$ and \H$ = 17.42\pm0.02$), this star turns out to be
about 0.5-1 magnitude fainter than the main sequence (MS) turn off
(TO) and  bluer than the MS both in the (\U, \U-\V) and in the
(\V, \V-\I) color magnitude diagrams (CMDs; see Figure \ref{cmd}).
Even more interesting is the comparison of the photometric properties
among the three epochs of observations. Unlike the CMD location in
EP2, the magnitudes derived for EP1 (\VV$=21.17\pm0.06$ and
\UU$=23.04\pm0.21$) and for EP3 (\B$=22.50\pm0.03$, \R$=20.60\pm0.03$
and \Ha$=20.27\pm0.03$)  approximately locate the star onto the MS.
Unfortunately, given the different instruments and filters, it is not
possible to directly compare the magnitudes but, both from the visual
inspection of images (see Figure \ref{map}) and from the CMD locations
with respect to the TO point, it turns out that during EP1 and EP3 the
star was about 2-3 magnitudes fainter than the TO, and hence $\sim 2$
mag fainter than in EP2.  This likely indicates that the observations
during EP1 and EP3 sampled the object in quiescence, while EP2 data
caught the star in an outburst state.  In addition, during each epoch
a magnitude modulation is present, with an indication of a smaller
amplitude in EP3 with respect to the variability detected during the
EP2 outbursting state. In fact, the frame-to-frame magnitude scatter
of the peculiar star during the outburst epoch (EP2) is $10-20\sigma$
larger than the scatter of normal stars in the same magnitude bin,
while this value decreases to $\sim4\sigma$ in EP3.

\begin{figure*}
\includegraphics[width=150mm]{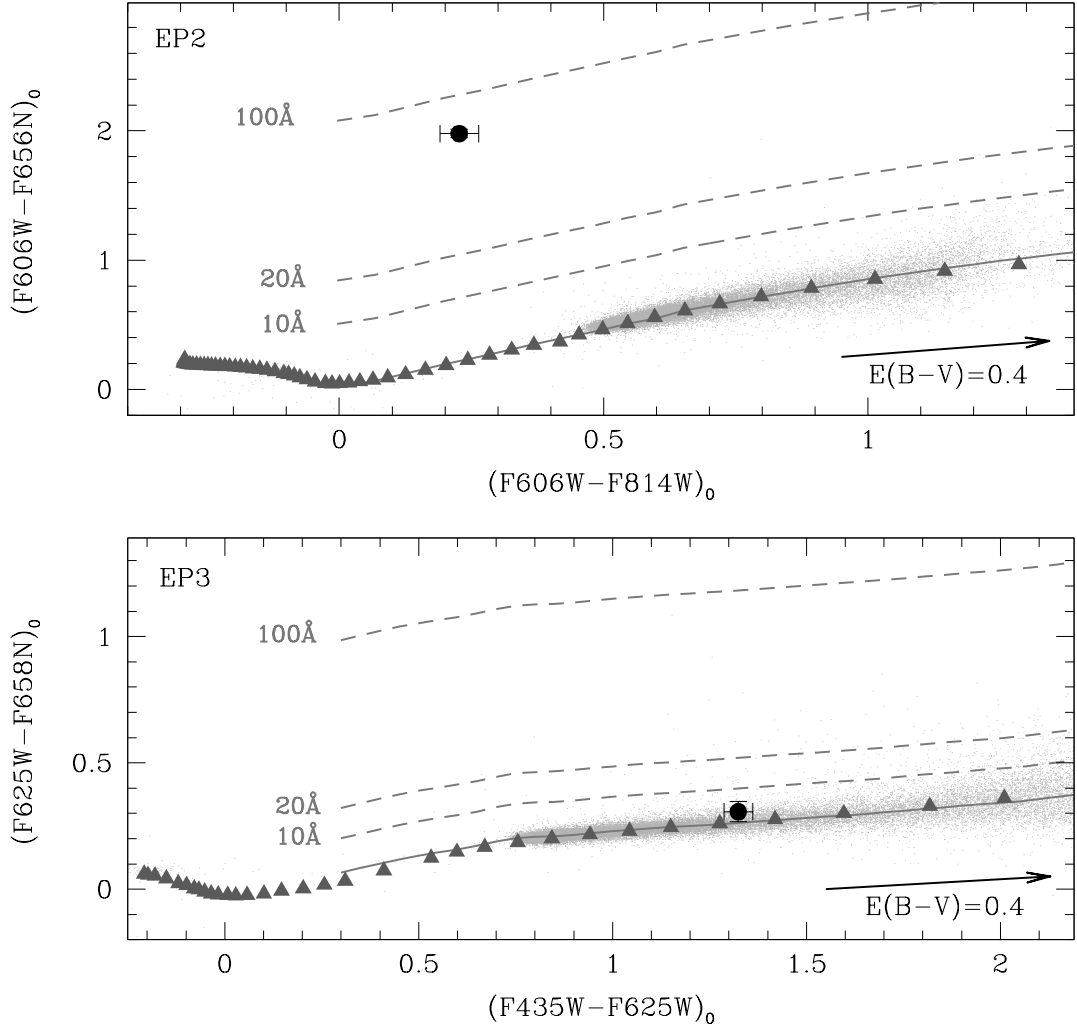}
\caption{Reddening corrected color-color diagrams for both EP2 and EP3. In each panel the solid line is the median color of stars (gray dots) with no \halpha\  excess  and hence the location of stars with EW$_{{\rm H\alpha}}=0$. It well corresponds to the location (gray triangles) predicted from  atmospheric models \citep{bessel98}. Dashed lines show the expected position for stars with increasing levels of \halpha\  emission, with the corresponding  EW$_{\rm H\alpha}$ labelled. The black dots mark the positions of the optical counterpart to \igr\ in each epoch. During the outburst (upper panel) its H$\alpha$ emission corresponds to  EW$_{\rm H\alpha}=71.6^{+5.5}_{-5.1}$ \AA, while in the quiescent state (lower panel) the star is located on the continuum reference line of stars with no  \halpha\ excess.  }
\label{halpha}
\end{figure*}

 In principle, for actively accreting LMXBs,  \halpha\ emission is expected 
from the accretion disk, while the contribution from the heated 
companion star should be minimal or even absent.
A visual inspection of EP2 images already suggests that this peculiar
star also has \halpha\ excess: in fact, in the \H\ image (panel $(e)$
in Figure \ref{map}) it is significantly brighter than its southern
neighbor, while these two objects show essentially the same magnitude
in broad band filters (as the \U, see panel $(b)$ in Figure
\ref{map}).   In order to quantify this excess we used a photometric technique 
based on the comparison between the magnitudes obtained from broad band  and  \halpha\ narrow filters 
\citep[][]{cool95}. In particular, in this work 
we used a method 
commonly applied to star forming regions (De Marchi et al., 2010) and
recently tested for the first time in the GC 47Tucanae \citep[Beccari et
al., MNRAS submitted;  see also][]{be13}. First of all, we corrected all magnitudes for
reddening by adopting $E(B-V)=0.4$ \citep{harris96}, then we selected
the peculiar star in the (\V-\H)$_0$ vs (\V-\I)$_0$ color-color
diagram.  Note that this color combination well samples the continuum
of stars with no \halpha\ emission for different spectral types
through the (\V-\I)$_0$ color index, and it provides a good estimate
of the \halpha\ emission through the (\V-\H)$_0$ color index, since
the \halpha\ line contribution to the \V\ band is negligible.  The
\halpha\ excess ($\Delta$\halpha) can be evaluated from the distance
between the (\V-\H)$_0$ color index of the considered star and an
empirical line\footnote{The reference line for the continuum has been
  determined from the median (\V-\H)$_0$ color as function of
  (\V-\I)$_0$ for stars with combined photometric error smaller than
  0.05 magnitudes. As shown in Figure \ref{halpha}, this empirical
  relation agrees very well with the theoretical one, obtained from
  atmospheric models \citep{bessel98}. We also emphasize that, while
  M28 may be affected by mild differential reddening, the reddening
  vector is almost parallel to the empirical line tracing the
  continuum (see Figure \ref{halpha}). This
  means that even large fluctuations in the reddening   of individual
  stars would not significantly affect the identification of objects
  with \halpha\ excess \citep{be10}. } representative of the
continuum.  In addition, the equivalent width (EW) of the
\halpha\ emission can be quantitatively estimated from
$\Delta$\halpha\ by applying equation (4) in \citet{demarchi10}: ${\rm
  EW}_{{\rm H\alpha}}={\rm RW} \times [1-10^{-0.4\times\Delta{\rm
      H}\alpha}]$, where RW is the rectangular width
of the filter,  \citep[see Table 4 in][]{demarchi10}.  With such a
method we estimated the \halpha\ excess ($\Delta {\rm H\alpha}=
1.98\pm0.03$;   upper panel in Figure \ref{halpha}) and the EW of the
\halpha\ emission (${\rm EW_{H\alpha}}=71.6^{+5.5}_{-5.1}$ \AA, where
the uncertainties take into account the errors in both colors) during
the EP2 outburst state.  By applying an analogous method to EP3 data,
making use of a suitable combination of \B, \R\ and \Ha\ filters, we find that
the star is located on the continuum reference line during its quiescent state 
(see the lower panel in Figure \ref{halpha}). Hence there is no
indication of \halpha\ emission in that epoch. 

Finally, we tried to investigate the possible presence of UV emission
by using the EP1 dataset in filters \nUV\ and \fUV. No source is
detected at the location of the peculiar star, most probably because
the images are not deep enough to reach its faint magnitudes.

\section{DISCUSSION AND CONCLUSIONS}\label{conclusions}
The photometric analysis revealed the presence of a very peculiar
star, which  underwent a strong luminosity increase and showed significant
\halpha\ excess in EP2.
 Even if this optical outburst occurred a few years before the INTEGRAL discovery, this evidence, combined with the
positional coincidence with the ATCA variable source recently detected
by Pavan et al. (Atel \#4981) and with the Chandra X-ray source \#23
revealed by \citet{becker03} and firmly associated to \igr\ by
Homan (ATel \#5045), strongly suggests that we have identified the
optical counterpart to \igr.
 Indeed several outbursts separated by a few years delay are quite 
typical of LMXBs containing a NS \citep[e.g 4U 1608-52, Aquila X-1;][]{asai12}.

Unlikely, the poor and irregular time coverage of our data prevented
us to firmly determine the period of the magnitude modulation, which
is expected to be correlated with the binary orbital motion.  However,
the non regular shape of the light curve (likely due to an
insufficient sampling of the orbital period), seems to suggest that the
variability is occurring in a timescale a few times shorter than the
duration of the observations ($\sim 5 - 10$ hours).  This seems to be
in agreement with the known properties of previously identified low
mass X-ray transients, that have orbital periods in the range between
40 min and 4.3 hours \citep{d'avanzo09} and even shorter in GCs
\citep[][and reference therein]{homer01, zurek09}.

During the quiescent state the companion star is approximately located on the MS,
$\sim 3$ magnitudes fainter than the TO, while during the outburst it
is $\sim 2$ magnitudes brighter and it is characterized by a bluer
color. As known from the study of companions to MSPs and LMXBs, such
an anomalous position is indicative of a perturbed state \citep[see,
  e.g.][]{fe01,cocozza08,pallanca10,testa12}.  In fact tidal deformations,
heating processes and the presence of an accretion disk can
significantly affect the magnitude and temperature of the star (E. Dalessandro et al., in preparation), thus
also altering its position in the CMDs.  The main tool to discriminate
between these effects is the determination of the light curve shape, but the
available data-sets prevent us to perform this study.

Finally, the presence of strong \halpha\ emission (with ${\rm EW}_{\rm
  H\alpha}=71.6^{+5.5}_{-5.1}$ \AA) during the outburst phase suggests
the presence of material accreting onto the NS.  On the other hand,  
no \halpha\ has been detected in quiescence, in agreement with the
 fact that when the accretion rate is slow, the disk is  much weaker and 
the \halpha\ absorption from the companion star dominates the spectrum.

Further optical studies are required to better constrain this system.
First of all, a photometric follow up, with a suitable time sampling
is needed to obtain accurate light curves and hence 
constrain the orbital parameters of the system.
Also a spectroscopic analysis, that, given the high crowding 
and the relative faint magnitude, is possible only during a  bright state, 
could help to characterize such system and the possible presence of an accretion disk through  the study  of the 
radial velocity curve, the chemical abundance patterns and UV emission lines.
 However to properly derive the companion radial velocity curve, it is needed to detect the spectral lines
associated with the companion  and to avoid those coming from the accretion disk. 

After the submission of this paper, XMM observations suggested that IGR J18245-2452 is the same source observed in the radio band as PSR J1824-2452I \citep[][arXiv:1305.3884v1]{papitto13}.

\section{Acknowledgement}
This research is part of the project COSMIC-LAB (www.cosmic-lab.eu)
funded by the European Research Council (under contract
ERC-2010-AdG-267675).  GB acknowledges the European Community's
Seventh Framework Programme under grant agreement no. 229517.

\end{document}